\documentclass{article}
\usepackage{bm}

\begin{document}

\title{Deriving the static interaction between electric dipoles via the quantum
gauge transformation} %Title of paper
%\author{Makoto Morinaga}
%\institute{
%Makoto Morinaga \at
%Institute for Laser Science, University
%of Electro-Communications, 1-5-1 Chofugaoka, Chofu, Tokyo, 182-8585, JAPAN\\
%\email{morinaga@ils.uec.ac.jp}}
\author{Makoto Morinaga
\thanks{Institute for Laser Science, University
of Electro-Communications, 1-5-1 Chofugaoka, Chofu, Tokyo, 182-8585, JAPAN
{e-mail:\it morinaga@ils.uec.ac.jp}
}}

%\date{\today}
\date{}

\maketitle %\maketitle must follow title, authors, abstract and \pacs

\begin{abstract}
Gauge transformation leaves the electric and the magnetic fields
unchanged as long as the gauge function is treated classically.
In this paper we consider the gauge transformation commonly
used to obtain the electric dipole interaction Hamiltonian in a system of
dipoles and the electromagnetic field
(G\"oppert-Mayer transformation) and treat the vector potential
that appear in the gauge function as an operator. While it
modifies the electric field,
the static interaction between the dipoles is derived.
\end{abstract}

% Body of paper goes here. Use proper sectioning commands. 
% References should be done using the \cite, \ref, and \label commands
\section{Introduction}
In the standard textbooks of quantum mechanics and electromagnetism,
A: the notion of gauge transformation is introduced before fields
are quantized so that the gauge function is treated classically, and
B: only time-varying fields are considered in the field quantization.
Derivation of the static force between charges or dipoles in the quantum regime,
i.e., as a result of the exchange of virtual photons
between them requires some amount of knowledges
on the quantum field theory\cite{az}. In this paper, we take up the
G\"oppert-Mayer transformation
and perform the transformation while treating the
vector potential that appear in this transformation as
an operator.
As a conseqence of this ``quantum'' transformation,
it is shown that the static interaction between the dipoles emerges and
the static dipole-field is added
to the electric field operator.

\section{G\"oppert-Mayer transformation}
Consider an atom consists of an electron of charge $-e$ orbiting
around an nucleus of charge $+e$ fixed at the origin.
The electron is also interacting with the external electromagnetic field
in addition to the Coulomb field created by the nucleus.
We choose Coulomb gauge ($\phi=0$) for the external electromagnetic field.
Then the Schr\"odinger equation for the electron is written as
\begin{equation}
 i\hbar\frac\partial{\partial t}|\varphi\rangle
=\left\{\frac 1{2m}[\bm p+e\bm A(\bm r,t)]^2
 +V(\bm r)\right\}|\varphi\rangle
\label{seqs0}
\end{equation}
where $V(\bm r)$ is the Coulomb potential created by the nucleus.
In the long wavelength approximation, the value of the vector potential
$\bm A(\bm r,t)$ is
approximated by $\bm A(\bm 0,t)$, so that (\ref{seqs0}) becomes
\begin{equation}
 i\hbar\frac\partial{\partial t}|\varphi\rangle
=\left\{\frac 1{2m}[\bm p+e\bm A(\bm 0,t)]^2
 +V(\bm r)\right\}|\varphi\rangle.
\label{seqs}
\end{equation}
The G\"oppert-Mayer transformation\cite{gm} is a gauge transformation
using the gauge function
\[
 \chi(\bm r,t)=-\bm r\cdot\bm A(\bm 0,t),
\]
that transforms the state $|\varphi\rangle$ to
$|\tilde\varphi\rangle=T|\varphi\rangle$ with 
a unitary transformation
$T=\exp(-i\frac e\hbar\chi(\bm r,t))
=\exp(-\frac i\hbar
\bm d\cdot\bm A(\bm 0,t))$
where $\bm d=-e\bm r$ is the dipole moment of the atom.
Substituting $|\varphi\rangle
=T^{-1}|\tilde\varphi\rangle$ into (\ref{seqs}),
or by first writing (\ref{seqs}) in the gauge independent form
and then using the relation
$\tilde{\bm A}(\bm r,t)=\bm A(\bm r,t)+\nabla\chi(\bm r,t)$
and $\tilde\phi(\bm r,t)=\phi(\bm r,t)-\partial_t\chi(\bm r,t)$,
an equation for $|\tilde\varphi\rangle$ is obtained:
\begin{equation}
 i\hbar\frac\partial{\partial t}|\tilde\varphi\rangle
=\left\{\frac 1{2m}\bm p^2
 +V(\bm r)-\bm d\cdot\bm E(\bm 0,t)
\right\}|\tilde\varphi\rangle.
\label{seqs2}
\end{equation}
Here we also used the relation $\bm E(\bm r,t)=-\partial_t\bm A(\bm r,t)$
applicable for the Coulomb gauge.
The G\"oppert-Mayer transformation has an aspect that
it rewrites the Hamiltonian into a standard form, i.e.,
as a sum of the kinetic energy and the potential energy.

\section{Transformation in the quantum regime}
In deriving (\ref{seqs2}), we implicitly assumed that $\partial_tT=
-i\frac e\hbar T\partial_t\chi(\bm r,t)$
which holds no longer if $\chi(\bm r,t)$ is an
operator-valued function.
Now we consider $n$ atoms
interacting with the electromagnetic field,
and perform the transformation in the quantum regime.
The Schr\"odinger equation in the long wavelength
approximation is written as
\begin{equation}
 i\hbar\frac\partial{\partial t}|\varphi\rangle
=\sum_q\left\{\frac 1{2m}[\bm p_q+e\bm A(\bm R_q,t)]^2
 +V(\bm r_q-\bm R_q)\right\}|\varphi\rangle
\label{seq}
\end{equation}
where $\bm r_q$ ($\bm R_q$) is the position of the
electron (nucleus) of the $q$th atom.
The unitary transformation $T$ on the state as
$|\tilde\varphi\rangle=T|\varphi\rangle$ is now given by
$T=\exp(-\frac i\hbar
\sum_q\bm d_q\cdot\bm A(\bm R_q,t))$
where $\bm d_q=-e(\bm r_q-\bm R_q)$ is the dipole moment of the
$q$th atom.
We shall rewrite (\ref{seq}) in terms of
$|\tilde\varphi\rangle=T|\varphi\rangle$ and $\tilde{\bm E}(\bm r,t)
=T\bm E(\bm r,t)T^\dag$ while treating $\bm A(\bm r,t)$
as an operator
(if $\bm A(\bm r,t)$ is treated as a classical quantity, then
we obtain just a set of $n$ equations each equivalent to (\ref{seqs2})).
It is readily shown that $|\tilde\varphi\rangle$ obeys
\begin{equation}
 i\hbar\frac\partial{\partial t}|\tilde\varphi\rangle
=\left\{\sum_q\left[\frac 1{2m}\bm p_q^2
 +V(\bm r_q-\bm R_q)\right]-i\hbar T(\partial_tT^{-1})
\right\}|\tilde\varphi\rangle.
\label{seq3}
\end{equation}
Writing $T$ as $T=e^X$ with $X=-\frac i\hbar
\sum_q\bm d_q\cdot\bm A(\bm R_q,t)$ and going back to
the definition of the exponential
$e^X=\lim_{m\rightarrow\infty}\left(1+\frac Xm\right)^m$,
it is shown that $T\partial_tT^{-1}$ can
be written generally as
\[
 T\partial_tT^{-1}=\int_0^1 ds\,e^{sX}Ye^{-sX}
 =\int_0^1 ds\,T_sYT_{-s}
\]
where $Y=\partial_tX=\frac i\hbar
\sum_q\bm d_q\cdot\bm E(\bm R_q,t)$ and $T_s=e^{sX}$.
Define a function $f(s)$ as $f(s)=T_sYT_{-s}$.
Then
\begin{equation}
 f'(s)=T_s[X,Y]T_{-s}=[X,Y]
\label{Z}
\end{equation}
using the fact that $[X,Y]$ commutes
with $X$ in our system as will be shown below. Solving
(\ref{Z}) as
\begin{equation}
 T_sYT_{-s}=f(s)=Y+s[X,Y]
\label{TT}
\end{equation}
which leads to
\[
 T\partial_tT^{-1}=\int_0^1 ds\,f(s)=Y+\frac 12[X,Y]
 =T_{\frac 12}YT_{-\frac 12}=\tilde Y-\frac 12[X,Y]
\]
with $\tilde Y=TYT^{-1}$. Now we expand $[X,Y]$:
\[
 [X,Y]=\frac 1{\hbar^2}\sum_{qq'}\sum_{m\,m'}d_{q\,m}d_{q'm'}
 [A_m(\bm R_q,t),E_{m'}(\bm R_{q'},t)].
\]
The general commutation relation between the vector potential operator and the
electric field operator is given by (24) and (27) of COMPLEMENT $\rm{C_{III}}$
of \cite{pa} (p.226-7).
From this, the simultaneous commutation relation is calculated as
\begin{equation}
 [A_m(\bm R,t),E_{m'}(\bm R',t)]=
 \frac{i\hbar}{4\pi\epsilon_0}
 \frac 1{\rho^3}\left(\delta_{mm'}-3\,\hat\rho_m\hat\rho_{m'}\right)
\label{cr}
\end{equation}
where $\bm\rho=\bm R-\bm R'$, $\rho=|\bm\rho|$,
and $\hat{\bm\rho}=\frac{\bm\rho}\rho$
(we omit here the term proportional to $\delta(\bm\rho)$).
Note that the right-hand side of the above equation is a $c$-number.
Finally putting all togther into (\ref{seq3}), we obtain
\begin{equation}
 i\hbar\frac\partial{\partial t}|\tilde\varphi\rangle
=\left\{H_0+H_{ext}
+\sum_{q>q'}\epsilon_{dip}(\bm R_q-\bm R_{q'},\bm d_q,\bm d_{q'})
+\epsilon_{self}\right\}|\tilde\varphi\rangle,
\label{seq4}
\end{equation}
with $H_0=\sum_q\left\{\frac 1{2m}\bm p_q^2+V(\bm r_q-\bm R_q)\right\}$,
$H_{ext}=-\sum_q\bm d_q\cdot\tilde{\bm E}(\bm R_q,t)$,
\[
 \epsilon_{dip}(\bm R,\bm d,\bm d')
=\frac 1{4\pi\epsilon_0}\frac 1{R^3}
\{\bm d\cdot\bm d'-3(\bm d\cdot\hat{\bm R})(\bm d'\cdot\hat{\bm R})\},
\]
and $\epsilon_{self}=\frac 12\sum_q\epsilon_{dip}(\bm 0,\bm d_q,\bm d_q)$
is the dipole self energy which is singular.
The $\epsilon_{dip}$ terms describe the static interaction
between different dipoles where as $H_{ext}$ corresponds to the
electric dipole interaction with the external field.
Redefining $Y$ as $Y=\bm E(\bm R,t)$ and using the
relation (\ref{TT}), $\tilde{\bm E}(\bm R,t)$ is calculated as
\[
 \tilde{\bm E}(\bm R,t)=TYT^\dag=T_1YT_{-1}=Y+[X,Y].
\]
Again using the commutation relation (\ref{cr}),
the relation between the electric field operators before and after the gauge
transformation is calculated as
\[
 \bm E(\bm R,t)=
 \tilde{\bm E}(\bm R,t)+\sum_q\bm E_{dip}(\bm R-\bm R_q,\bm d_q)
\]
where
\begin{equation}
 \bm E_{dip}(\bm R,\bm d)=
 -\frac 1{4\pi\epsilon_0}
 \frac 1{R^3}\left\{\bm d-3(\bm d\cdot\hat{\bm R})\hat{\bm R}\right\}
\label{dip}
\end{equation}
is the electric field created by a dipole of dipole moment $\bm d$
at the distance $\bm R$.
We see that the electric field operators in the two pictures differ by
the electric field created by $n$ dipoles.

\section{Conclusion and discussion}
By treating the gauge function of the G\"oppert-Mayer transformation
as an operator, the static interaction between the dipoles is derived.
Let us call the picture before (after) the transformation
P1 (P2). It is noted that, in P2,
the static dipole interaction term is present even in the absence
of photons.
Going back to P1, the corresponding state
contains (virtual) photons which yield forces between the dipoles.

\appendix
\section{Coulomb potential}
In this appendix we look for a unitary transformation $T$ which leads to
the Coulomb potential:
\[
 \bm E(\bm r,t)=\tilde{\bm E}(\bm r,t)+\bm E_c(\bm r)
\]
where $\tilde{\bm E}(\bm r,t)=T\bm E(\bm r,t)T^\dag$ and
$\bm E_c(\bm r)=\frac q{4\pi\epsilon_0}\frac{\bm
r}{r^3}=-\nabla\phi_c(r)$ is the electric field derived from the
Coulomb potential
$\phi_c(r)=\frac q{4\pi\epsilon_0}\frac 1r$.
The relation
\[
\begin{array}{rl}
 \bm E_{dip}(\bm r-\bm s,-q\,d\bm s)=&\frac q{4\pi\epsilon_0}\frac 1
{|\bm r-\bm s|^3}
\left\{d\bm s-3\frac{[d\bm s\cdot(\bm r-\bm s)](\bm r-\bm s)}
{|\bm r-\bm s|^2}\right\}\\
=&-(d\bm s\cdot\nabla_{\bm s})\bm E_c(\bm r-\bm s)
\end{array}
\]
(see (\ref{dip})) suggests to use
$T=\exp(\frac i\hbar
q\int_{\bm 0}\bm A(\bm s,t)\cdot d\bm s)$ (instead of
$T=\exp(-\frac i\hbar\bm d\cdot\bm A(\bm 0,t))$
of the G\"oppert-Mayer transformation)
where the integration path goes from the origin to somewhere
infinity (imagine a chain of dipoles that connects the origin
to infinity).
$T$ itself depends on the integration path, but $\tilde{\bm E}$
does not as we shall see.
Again writing $T$ as $T=e^X$ with $X=\frac i\hbar
q\int_{\bm 0}\bm A(\bm s,t)\cdot d\bm s$
and defining $Y$ as $Y=E_m(\bm r,t)$,
then (\ref{TT}) leads to
\begin{equation}
 \tilde E_m(\bm r,t)=f(1)=X+[X,Y].
\label{TET}
\end{equation}
Using the simultaneous commutation relation (\ref{cr})
in the form
\begin{equation}
\begin{array}{rl}
 [A_{m'}(\bm s,t),E_m(\bm r,t)]=&\frac{i\hbar}{4\pi\epsilon_0}
 \frac 1{\rho^3}\left(\delta_{mm'}-3\,\hat\rho_m\hat\rho_{m'}\right)\\
=& \frac{i\hbar}{4\pi\epsilon_0}
\frac\partial{\partial\rho_{m'}}\frac{\rho_m}{\rho^3}\\
%=& -\frac{i\hbar}{4\pi\epsilon_0}
%\frac\partial{\partial\rho_m}\frac\partial{\partial\rho_{m'}}
%\frac 1\rho
\end{array}
\label{cr_}
\end{equation}
where $\bm\rho=\bm s-\bm r$, $\rho=|\bm\rho|$,
and $\hat{\bm\rho}=\frac{\bm\rho}\rho$, $[X,Y]$ is calculated as
\[
\begin{array}{rl}
 [X,Y]&=\frac i\hbar q\int_{\bm 0}\sum_{m'}
 [A_{m'}(\bm s,t),E_m(\bm r,t)]ds_{m'}\\
 &=\frac i\hbar q\int_{\bm 0}\sum_{m'}
\frac{i\hbar}{4\pi\epsilon_0}
\frac\partial{\partial s_{m'}}\frac{s_m-r_m}{|\bm s-\bm r|^3}ds_{m'}\\
&=-\frac q{4\pi\epsilon_0}
\int_{\bm 0}d\bm s\cdot\nabla_{\bm s}\frac{s_m-r_m}{|\bm s-\bm r|^3}\\
&=-\frac q{4\pi\epsilon_0}\frac{r_m}{r^3}.
\end{array}
\]
Substituting this into (\ref{TET}) we obtain
\[
 \tilde{\bm E}(\bm r,t)=\bm E(\bm r,t)
-\frac q{4\pi\epsilon_0}\frac{\bm r}{r^3}
\]
as expected.


\begin{thebibliography}{9}
 \bibitem{az}  Zee, A.:
Quantum Field Theory in a Nutshell,
Princeton University Press (2003)
 \bibitem{gm} G\"oppert-Mayer, G., Ann. Phys., 401,
273 (1931)
 \bibitem{pa} Cohen-Tannoudji, C., Dupont-Roc, J., Grynberg, G.:
Photons and Atoms, John Wiley and Sons (1987)
\end{thebibliography}
\end{document}